\documentclass[runningheads]{llncs}
\usepackage[T1]{fontenc}
\usepackage{graphicx}
\usepackage{amsmath}
\usepackage{bussproofs}
\usepackage{algorithm}
\usepackage{algpseudocode}
\usepackage{svg}
\usepackage{enumitem}
\usepackage{mathptmx}
\usepackage{stmaryrd}
\usepackage{amssymb}
\usepackage{tikz}
\usetikzlibrary{automata,positioning}
\usepackage[
  subtle
]{savetrees}

\usepackage{verbatim}
\usepackage{listings}
\newenvironment{allintypewriter}%
  {\endgraf\ttfamily\verbatim}%
  {\endverbatim}

\begin{document}
\title{BinSub: The Simple Essence of Polymorphic Type Inference for Machine Code}
\author{Ian Smith\orcidID{0009-0006-5584-9345}} 
\authorrunning{Ian Smith}

\institute{
Trail of Bits, New York, USA\\
\email{ian.smith@trailofbits.com}}

\maketitle              % typeset the header of the contribution
\begin{abstract}
Recovering high-level type information in binaries is a key task in reverse engineering and binary analysis. Binaries contain very little explicit type information. The structure of binary code is incredibly flexible allowing for ad-hoc subtyping and polymorphism. Prior work has shown that precise type inference on binary code requires expressive subtyping and polymorphism.

Implementations of these type system features in a binary type inference algorithm have thus-far been too inefficient to achieve widespread adoption. Recent advances in traditional type inference have achieved simple and efficient principal type inference in an ML like language with subtyping and polymorphism through the framework of algebraic subtyping. BinSub, a new binary type inference algorithm, recognizes the connection between algebraic subtyping and the type system features required to analyze binaries effectively. Using this connection, BinSub achieves simple, precise, and efficient binary type inference. We show that BinSub maintains a similar precision to prior work, while achieving a 63x improvement in average runtime for 1568 functions. We also present a formalization of BinSub and show that BinSub's type system maintains the expressiveness of prior work.

\keywords{Type inference \and subtyping \and algebra \and reverse engineering \and binary analysis.}
\end{abstract}
%
%
%

% Snippet for adding SAS badges
%\begin{center}
%  \includegraphics[scale=0.16]{ValidatedBadge.png} \includegraphics[scale=0.16]{ExtensibleBadge.png} %\includegraphics[scale=0.16]{AvailableBadge.png}%
%\end{center}

\section{Introduction}
Binary reverse engineering has a variety of important security applications including binary hardening, vulnerability research, program understanding, and recompilation. Decompilers are widely used to aid in the reverse engineering process, providing high-level understandable representations to downstream applications. Type information drives the quality of decompilation. Types determine how pointer arithmetic is rewritten to field or array accesses, the form of function calls, and what casts are required \cite{scalbinaryrewriting,Enders_2023}. The correctness of this type information has a direct effect on the decompilation's usability for vulnerability assessment.

While expensive and precise methods exist for type inference on binaries, widely used decompilers take simple approaches to recovering type information. Ghidra and IDA both by default provide type inference passes that directly propagate type equality from a called function to its callers through identified parameters' reaching definitions \cite{simpleprop,skochinksy_igors_2022,noauthor_ghidra_nodate}. This approach works well on binaries with significant type information, safely propagating known function types to callers. The approach is simple to understand and implement, and scales to large binaries with acceptable precision. This approach unfortunately has major shortcomings when dealing with novel binaries that do not have type library information to bootstrap analysis. Even in binaries that use a significant number of well known libraries, new types created by the target application are left entirely to the reverse engineer to recover manually. Additionally, type inference based on equality between parameters will under refine a pointer that has capabilities (e.g., fields) that are only used by some callees of the function. Direct equality ignores complexities like width subtyping \cite{tiebap}. A fully stripped and novel binary confounds current analyses applied in these disassemblers \cite{tiebap,retypd}.

New work on type systems for binary type inference have attempted to address these limitations by adding features to the type system to cope with the flexibility of C and binary types. Early approaches applied a purely unification based type inference algorithm, lacking subtyping constraints between variables \cite{10.1007/978-3-540-69738-1_1}. This approach can drastically over-refine type solutions for type variables by assuming type dataflow is symmetric. TIE refined this approach with subtyping both for atomic types and depth and width subtyping for records \cite{tiebap}. Retypd, achieved further precision improvements by adding polymorphism, contravariant stores, and recursive types \cite{retypd}. 

Unfortunately, these more expressive type systems that compensate for the flexibility of binary code have resulted in complex constraint simplification algorithms. Retypd's simplification algorithm for recursively constrained polymorphic types translates type constraints to a weighted pushdown system that recognizes legal derivations between interesting variables. Derivations are then explored through a saturation procedure that is $O(N^3)$ in the size of the type constraints \cite{caucal1992regular}. The algorithm is inefficient on binaries containing large functions or functions with many callees (due to callsite instantiation). Furthermore, the algorithm contains significant subtleties, particularly related to width subtyping, cycle elimination and path exploration, and the extensions to weighted pushdown automata saturation \cite{peter_aldous_retypd_nodate}. The complexity of the analysis limits the application of the type system in more widely used binary analysis platforms.

The contribution of this paper is to show that Retypd's type system can be reformulated within the framework of algebraic subtyping, resulting in a 63x average speedup on an evaluation dataset of 1568 functions \cite{algebraicdolan}. This reformulation leads to a simple type system that preserves the expressive features of Retypd: subtyping, contravariance of pointer stores, recursive types, and polymorphism. Placing binary type inference in the context of algebraic subtyping challenges the problem's pretensions as a form of type inference outside of traditional type inference. Consequently we show that we are justified in demanding similar performance and advances for binary type inference as those seen in high-level programming language type systems. Specifically we:
\begin{itemize}[noitemsep,topsep=0pt]
    \item demonstrate that Retypd's type system features can be replicated simply and efficiently using the framework of algebraic subtyping
    \item highlight a practical implementation of this type inference algorithm in the Angr binary analysis framework: BinSub
    \item empirically evaluate BinSub against Retypd, demonstrating similar precision with marked performance improvements (63x reduction in average runtime on 1568 functions sampled from the ALLSTAR armel and amd64 categories \cite{allstarDataset})
    \item present a formalization of BinSub and demonstrate a translation between Retypd constraints and BinSub constraints along with a preservation of subtyping judgement derivations 
\end{itemize}

\section{Background}

\subsection{Binary Type Inference}\label{sec:bintygen}

 Approaches to static type recovery can roughly be separated into two phases: constraint generation and constraint solving. Binary type inference algorithms typically take similar approaches to constraint generation. A model of a binary's dataflow is built through a collection of static analyses that identify function boundaries, resolve function calls, determine function parameter locations based on the target application binary interface (ABI), and perform pointer analysis to propagate dataflow through loads and stores  \cite{tiebap,10.1007/978-3-540-69738-1_1,scalbinaryrewriting,retypd}. These analyses are often constructed as a static dataflow problem, using abstract interpretation frameworks such as value-set-analysis (VSA) \cite{balakrishnan2004analyzing}. Type constraints are then generated by visiting the program representation and generating a type constraint for each dataflow between program variables (i.e., registers, parameters, and recovered stack variables). Approaches then differ in the underlying type system features they support. 

We first motivate BinSub's simplifications to Retypd's type system by outlining the challenges the type system was designed to solve \cite{retypd}. We identify four pillars of type inference in Retypd: subtyping, pointer covariance/contravariance, recursive types, and polymorphism. 

\subsubsection{Subtyping:}\label{sec:subty}
Retypd first addresses a problem highlighted in TIE \cite{tiebap}. Compilers erase types, allowing fortuitous reuse of variable locations for values of different types as long as the type is small enough to use the location. Retypd highlights cases where a value can be used both as an integer and as a pointer when passed to a function with the signature \texttt{f(void*, int)}

\begin{figure}[H]
    \centering
    \begin{allintypewriter}
        xor eax, eax
        push eax
        push eax
        call f
    \end{allintypewriter}
    \caption{A function call using a zeroed register both as a null pointer and an integer}
    \label{fig:reg_reuse}
\end{figure}

Figure \ref{fig:reg_reuse} shows optimized compilation of the call \texttt{f(NULL, 0)}. The compiler is able to reuse the zeroed value of \texttt{eax} for both parameters. Unifying the types of the two parameters based on their equality to \texttt{eax} would cause both parameters of \texttt{f} to have the imprecise type $\top$ (assuming the supertype of \texttt{int} and \texttt{void*} is $\top$ and not a type error). This challenge extends to type punning on \texttt{void*} or union types. Binary type systems incorporate subtyping to avoid this imprecision. An assignment of the form \texttt{a=b} does not imply that the type of \texttt{a} is equal to that of \texttt{b}. Instead, for type variables $\alpha$ and $\beta$ for program variables \texttt{a} and \texttt{b}, constraints contain directionality and are of the form $\beta \leq \alpha$. Adding subtyping to a type system with type inference creates difficulties in presenting types to the user. Unification creates canonical types for each solved type. On the other hand, a type system based on subtyping collects a set of constraints for each type variable. Typically these constraints are simplified before presentation. Retypd performs simplification both by reducing constraint sets to relationships between interesting type variables and through automata minimization of type sketches. 

\subsubsection{Recursive Types:}\label{sec:rectype}

Retypd departs from TIE by handling recursive types in combination with subtyping. C-like languages often use recursive structure types to represent lists or trees. Type recovery algorithms targeting binaries have often struggled to construct these types \cite{10.1145/2896499}.

\begin{figure}[H]
    \centering
    \begin{allintypewriter}
        struct lst {
            int value;
            struct lst* next;
        };
    \end{allintypewriter}
    \caption{A recursive linked list type in C}
    \label{fig:linkedlist}
\end{figure}

Types like those presented in Figure \ref{fig:linkedlist} create problems for a naive type simplification algorithm. Specifically, constraints on a variable may reference the variable itself so naively applying typing rules to a fixed point will diverge. Retypd represents rules within an unconstrained pushdown system that is saturated to represent reachable derivations in the form of an automata. This process avoids diverging in the presence of recursive constraints. The automata may contain loops, indicating a recursive type. At this point the algorithm can cut loops by introducing a new type variable for strongly connected component (SCC) members (thus creating a recursive type at this new type variable).   

\subsubsection{Pointer Representation:}\label{sec:pointerrepproblem}

Subtyping between pointers presents further complications. Pointers can both be written to and read from. If a pointer is represented by a type constructor \texttt{ptr($\alpha$)} and that constructor is treated as covariant (as occurs in TIE) then a loss of type information occurs after a pointer has been copied from an old pointer $p$ to a new pointer $q$, implying the constraint $p \leq q$ \cite{tiebap}. If a value is stored to $q$, then loading from $p$ does not result in subtyping information being propagated between the store and the load due to the directionality of covariance on $p$ and $q$ \cite{retypd}. The parameter of the pointer $q$ is not a subtype of $p$ because the parameter is covariant. On the other hand, if the pointer constructor is treated as contravariant then storing into the old pointer $p$ will not be reflected in loads from $q$, because now the parameter of $p$ is not a subtype of the parameter of $q$. Intuitively, we would like subtyping to be implied both in the first and second case. This problem is exactly analogous to mutable \texttt{ref} types in ML, where the typical treatment when adding subtyping is to make mutable references invariant in their type parameter \cite{mlsub}. Such a solution is unacceptable when using type inference to recover types in a binary. The natural solution is to treat store and load as separate type parameters, where the load parameter is covariant and the store parameter is contravariant \cite{mlsub}. Retypd takes a similar approach and adds the following typing rule to ensure propagation between load and store capabilities: 

\begin{figure}[H]
    \centering
    \begin{prooftree}
        \AxiomC{Var $\alpha$.load}
        \AxiomC{Var $\alpha$.store}
        \BinaryInfC{$\alpha$.store $\leq$ $\alpha$.load}
    \end{prooftree}
    \caption{S-Pointer rule in retypd} 
    \label{fig:spointerrule}
\end{figure}

This rule creates the binding between the store and load capability for a type variable, allowing derivations that determine that a value stored to a pointer must be a subtype of the type loaded from that pointer. Unfortunately, this rule would result in an unbounded number of transitions in the unconstrained pushdown system if it were applied whenever it is possible to do so \cite{retypd}. Instead this rule is instantiated lazily during saturation by recognizing when a load or store is in the reaching push set of a node and adding a shortcut edge to the implied target of the S-Pointer rule. Retyd takes advantage of the strategy applied to handle recursive types to also handle S-Pointer rule instantiations without diverging.   

\subsubsection{Polymorphism:}\label{sec:polysec}

Finally, Retypd augments binary type inference with soundness in the presence of polymorphic functions. In C, \texttt{void*} types can be used to replicate polymorphsim, where a function performs an operation, agnostic to the underlying type of the parameter or return value. Standard heap allocation and free functions are examples of this behavior where there is no subtyping relationship between pointers returned from \texttt{malloc} or passed to \texttt{free}. If every type returned from \texttt{malloc} was placed in a subtyping relationship with $\alpha$ such that \texttt{malloc} is assigned the type $\texttt{size\_t} \rightarrow \alpha$, every heap allocated pointer in the program would be over-refined with capabilities from every other heap pointer. To resolve this issue Retypd adds polymorphism to the type system, allowing \texttt{malloc} to be given the type $\forall \alpha. \texttt{size\_t} \rightarrow \alpha$, generalizing the return type and avoiding over-refined types where malloc is used. Retypd defines type sketches which represent the derived constraints on a type variable as an automaton. Type sketches form a lattice. This lattice structure allows for joining type constraints for all uses of a given function to collect the most general type implied by all callsites of the function. Type sketches can be instantiated as constraints in order to use the sketch as part of the constraint set at each callsite. 

Supporting type inference directly for a type system with the features highlighted in section \ref{sec:bintygen} in Retypd resulted in a complex and difficult to analyze algorithm that relies on saturating a weighted pushdown system in order to produce a finite state transducer that recognizes legal derivations, a sketch inference algorithm, polymorphic schemes based on type sketches, and a type inference algorithm based on these components \cite{retypd}. The algorithm is computationally expensive and difficult to implement. In the following sections we show that Retypd's pillars of type inference can be captured in the framework of algebraic subtyping \cite{simplsub}, leading directly to an efficient and simple type inference algorithm for machine code: BinSub.   

\subsection{Algebraic Subtyping}

Algebraic subtyping, introduced in MLSub, formulates a subtyping lattice of ML types that has convenient algebraic properties that enable principal type inference of compact types in ML with depth and width subtyping for records \cite{algebraicdolan}. Algebraic subtyping creates a distributive lattice through the use of intersection and union types. Type inference restricts the represented types to polar types where intersection types only appear in negative positions (also referred to as input positions) and union types only appear in positive positions (also referred to as output positions). This syntactic restriction on types allows for a finite representation of types that suffices for principal type inference. Polar types also enable the construction of type automata used for type simplification. MLSub uses a modified version of Algorithm W: bi-unification, to produce polar types from subtyping constraints \cite{algebraicdolan}. Bi-unification performs substitutions separately on positive and negative occurrences of variables, replacing a negative occurrence with a variable's upper bounds and a positive occurrence with a variable's lower bounds. Polar types allow bi-unification to eliminate subtyping constraints by aggregating bounds on a variable through intersection and union types.

SimpleSub presents a simplified algorithm for bi-unification that is efficient \cite{simplsub}. The algorithm proceeds by first aggregating all implied subtyping constraints on a variable, and then coalescing the type of a variable by substituting negative and positive occurrences accordingly.

\section{Implementing BinSub}

BinSub's core idea is that the type system features of ML with subtyping correlate closely with the features required for precise binary type inference. Algebraic subtyping type systems like MLSub naturally support subtyping, recursive types, and parametric polymorphism with principal types  \cite{retypd,mlsub}. Consequently algebraic subtyping is a natural solution to binary type inference.

\subsection{Overview}\label{sec:overview}

BinSub acts as the type inference algorithm within a decompiler or binary analysis framework, ingesting subtyping constraints of the form $\tau_1 \leq \tau_2$ and solving for C types that instantiate type variables of interest (e.g., function variables to recover function signatures). These subtyping constraints are generated by walking the intermediate representation (IR) of binary functions and producing type constraints implied by a given instruction. 

\subsubsection{Function Lifting and Constraint Generation:}
Operating over a function's lifted IR assumes that disassembly and control flow graph (CFG) recovery have already produced an analyzable CFG for each function \cite{x86disasm,secondwritelifting}.  While this process is similar to type constraint generation in a typical high-level language, there are additional unknowns that require analysis to derive. When generating constraints for IR statements: \texttt{*(sp+4)=a;b=load (sp+4)}, constraint generation must record that \texttt{sp+4} points to a stack variable \texttt{s$_1$} in order to aggregate type information for that variable.  This aliasing information implies constraints \texttt{a}$\leq$\texttt{s$_1$} and \texttt{s$_1$}$\leq$\texttt{b} and aggregates constraints on the stack variable \texttt{s$_1$}. 

To encode this aliasing and variable information into type constraints, typically binary analyses first compute a memory access analysis such as VSA \cite{10.1007/978-3-540-69738-1_1} to statically compute an over-approximation of pointers accessed at a given load or store and the offsets accessed from a given pointer before performing type recovery. These offsets are used to compute a set of non-overlapping stack variables and record field constraints on pointers. Alternative constraint generation analyses are possible, including using dynamic analysis to record memory access offsets at runtime \cite{rewardsdynamic}. In our implementation of BinSub, described in section \ref{angrimpl}, we use Angr's fast variable recovery analysis to generate type constraints. This analysis resolves loads and stores that access stack variables by using a VSA-like domain to track values relative to the stack pointer and recording load and store offsets that are performed on those values \cite{angr_variables_fast_ail}. These analyses allow subtyping constraints to operate over abstract variables rather than concrete registers and memory locations relative to the stack pointer. Figure \ref{fig:exampleccode} and \ref{fig:irreprexample} shows the translation from C code to an intermediate representation with recovered stack references. BinSub performs type inference on a representation similar to Figure \ref{fig:irreprexample}; we use this function as a running example to demonstrate BinSub in Section \ref{sec:methodsbinsub}.

\begin{figure}[H]
\begin{minipage}{.5\linewidth}
    \begin{allintypewriter}
        struct lst* x; // free var 1
        void* y; // free var 2
        struct lst* curr = x;
        struct lst* cpy = y
        while(curr != NULL) {
            cpy = curr;
            curr = cpy->next;
            cpy->value++;
        }
        ... // use cpy here
    \end{allintypewriter}
    \caption{Incrementing values in a linked list}
    \label{fig:exampleccode}
  
  \end{minipage}%
  \begin{minipage}{.5\linewidth}
    \begin{allintypewriter}
        block_1:
            1: stack_slot_1 = x;
            2: stack_slot_2 = y;

        block_2:
            3: stack_slot_2 = stack_slot_1;
            4: t1 = load [stack_slot_2 + 4], 4;
            5: stack_slot_1 = t1;
            6: t2 = load [stack_slot_2], 4;
            7: t3 = t2 + 1
            8: store [stack_slot_2], t3
    \end{allintypewriter}
    \caption{Intermediate representation of Figure \ref{fig:exampleccode} after variable recovery}
    \label{fig:irreprexample}
  \end{minipage}
\end{figure}

\subsubsection{Type Inference:}
To start type inference, the IR displayed in \ref{fig:irreprexample} is translated into subtyping constraints. After constraint extraction, BinSub solves for C types for type variables in three steps: bi-unification, simplification, and lowering. Bi-unification aggregates constraints on a given variable and produces a finite BinSub type for that variable. These types are conservative in that the constraints on the variable are fully represented by intersection and union types, however, these types will have redundant intermediate type variables and overly complex structure \cite{simplsub}. Simplification applies subtyping rules to aggregate records, function types, and atoms by merging intersection and union types and removing redundant type variables. Finally, type lowering converts BinSub types to C types by selecting a representative C type for each component of a BinSub type. In the following section, we describe bi-unification, type simplification, and lowering within BinSub. In Section \ref{sec:formalbinsub}, we present BinSub's subtyping rules and relate them to Retypd's subtyping rules to demonstrate that this implementation preserves the expressive type system features of Retypd.

\subsubsection{Limitations:}
The approach outlined above restricts the domain of BinSub to inferring types for binaries where IR of the form shown in Figure \ref{fig:irreprexample} can be recovered. Type inference operates over this IR and its precision is bounded by the precision of the recovered IR. This IR corresponds to a low-level untyped C-like representation with abstract variables. Complex features of binaries are known to complicate code recovery, disassembly, and lifting. Indirect jumps can prevent control flow recovery when using recursive descent disassembly, non-standard application binary interfaces (ABIs) may prevent parameter or stack recovery, and self-modifying code may not be lifted all \cite{disasoverview}. In order to apply type inference to functions that are not represented properly by a disassembler, a user may be required to make manual fixes (e.g., manually labeling control flow). 

Given that BinSub operates over a C-like representation and recovers C types, type system features such as generic or templated types, classes, and virtual functions are not represented in the type system. The IR is a monomorphic representation of functions with stores, loads, and basic expressions that correspond to the C language. The next section outlines the types that can be expressed in BinSub: records, functions, pointers and atomic types. While the algorithm itself is parameterized on an arbitrary atomic type lattice, we instantiate BinSub with a lattice available in Angr that contains basic primitive C integers in Section \ref{angrimpl} \cite{wang2017angr}.

Finally, lowering a more expressive type system to C types for presentation requires heuristics \cite{retypd}. BinSub can collect types that would typically be a type error: $\texttt{int16} \sqcap \texttt{ptr}(\alpha, \beta)$. As described in the following section this type would be the subtype of a 16 bit primitive integer and pointer, resulting in bottom (a type denoting that no subtype exists). BinSub can represent such a type directly whereas C types cannot. The heuristics in \ref{sec:typelowering} describes how pointers are prioritized over atomic types when aggregating a type solution for a variable as a C type.

\subsection{Applying Algebraic Types to Binary Type Inference}\label{sec:methodsbinsub}

Figure \ref{fig:binsubtypesyntax} describes the syntax of BinSub types where $n_x$ are metavariables ranging over natural numbers.

\begin{figure}[H]
    \centering
    $\tau ::= \{ (n_0, n_1) : \tau, ... (n_{m-1}, n_{m}): \tau\} | (n_0: \tau, ... n_p: \tau) \rightarrow (n_{p+1}: \tau, ... n_r: \tau) | \texttt{ptr}(\tau,\tau) | \alpha | \top | \bot| \tau \sqcup \tau | \tau \sqcap \tau | \mu\alpha.\tau|\rho$

    $\rho = \texttt{int64} |\texttt{int}|\texttt{float}|...$
    \caption{BinSub Types}
    \label{fig:binsubtypesyntax}
\end{figure}

BinSub's type system consists of records, functions, pointers, type variables, and atomic types. Records are partial functions of field capabilities to types. A field capability is a combination of an offset into the record (in bytes) and a size (in bits). Tracking field capabilities allows BinSub to define a lattice of record types in the presence of overlapping loads or stores to or from a record. Similarly, function types are flexible in the number of parameters and return values. Function types are represented as a partial function from parameter index to type and a partial function from return index to type. Departing from type systems like TIE \cite{tiebap}, pointers are constructed from two types: a contravariant store type and a covariant load type. This separation of pointer types addresses unsoundness in mixing mutable references with subtyping and intersection types \cite{unsoundintersectiontypes,mlsub}. Atomic types, defined by the syntactic class $\rho$ are any set of atoms equipped with a complete subtyping lattice. Throughout the text the atomic lattice is referred to as $A$ and is equipped with standard lattice join and meet operations: $\sqcup_A$ and $\sqcap_A$ as well as a top and bottom element: $\top_A$ and $\bot_A$. BinSub is parameterized over any lattice of atomic types. BinSub types also contain recursive types of the form $\mu\alpha.\tau$ and type variables $\alpha$. Viewed under SimpleSub's formalization, recursive types can be understood as equivalent under some unrolling substitution of the recursive type for $\alpha$ \cite{simplsub}. Subtyping introduces some complication to the presentation by requiring hypothetical subtyping judgements to demonstrate subtyping judgements between recursive types that are not equivalent by unrolling. BinSub represents a collection of constraints on a type variable by aggregating constraints on the variable into an intersection ($\tau \sqcap \tau$) or union ($\tau \sqcup \tau$) type of its bounds.

\subsubsection{Polarity Restrictions:}

For ease of manipulation and to represent recursive types finitely, BinSub separates the syntax of types into negative and positive types \cite{simplsub,algebraicdolan}. Positive types represent outputs and negative types represent inputs. Constraints represent dataflow between positive and negative types in the form $\alpha^+ \leq \beta^-$. The polarity of BinSub types matches SimpleSub and MLSub with function returns being positive positions, parameters being negative positions, and record fields being positive positions. Pointer types are negative in the store parameter and positive in the load parameter: $\texttt{ptr}(\tau^-, \tau^+)$. Union types can only appear in positive positions, and intersection types can only appear in negative positions. These simplifying assumptions enable a translation from polar types to type automata, allowing for simplification of types through automata simplification \cite{mlsub}. Prior work has shown that type inference using polar types is complete for type-able expressions \cite{algebraicdolan}. Throughout the rest of the paper we denote positive types as $\tau^+$ and negative types as $\tau^-$.

It is interesting to compare polarity to the construction of Retypd's weighted pushdown system. Retypd maintains a similar distinction between a variable appearing in the left hand side of a constraint or right hand side of a constraint. Each interesting variable is tagged with an L or R when emitting rules. For an interesting variable x, $x_R$ collects lower bound on $x$, while $x_L$ collects upper bounds. In algebraic subtyping, variables are treated differently depending on if they appear positively or negatively in a constraint. A positive occurrence of $x$ collects lower bounds on a variable corresponding to $x_R$ and a negative occurrence collects upper bounds on a variable corresponding to $x_L$.
\subsubsection{Constraint Generation:}

Type inference operates over a set of constraints of the form $\tau^+ \leq \tau^-$. Section \ref{angrimpl} describes how these constraints are generated during Angr's variable recovery analysis.

\begin{figure}[H]
    \centering
    \begin{lstlisting}
        1: x$\leq$stack_slot_1
        2: y$\leq$stack_slot_2
        3: stack_slot_1$\leq$stack_slot_2
        4$_0$: stack_slot_2$\leq$ptr(a, {(4,4): b})$\wedge$a$\leq${(4,4): b}
        4$_1$: b$\leq$t1
        5: t1$\leq$stack_slot_1
        6$_0$: stack_slot_2$\leq$ptr(c, {(0,4): d})$\wedge$c$\leq${(0,4): d}
        6$_1$: d$\leq$t2
        7$_0$: t2$\leq$int32
        7$_1$: int32$\leq$t3
        8$_0$: stack_slot_2$\leq$ptr({(0,4): e}, f)$\wedge${(0,4): e}$\leq$f 
        8$_1$: t3$\leq$e
    \end{lstlisting}
    \caption{Constraints generated from Figure \ref{fig:irreprexample}}
    \label{fig:exampleconstraints}
\end{figure}

Figure \ref{fig:exampleconstraints} highlights constraints generated from the example linked list code. The subtyping constraints are for the most part straightforward, encoding the dataflow of the intermediate representation. When \texttt{stack\_slot\_2} is used as a pointer, that capability is enforced by requiring that \texttt{stack\_slot\_2} be a subtype of a pointer with fresh variables where applicable. Each instantiation of the \texttt{ptr} constructor also results in a side condition, enforcing subtyping between the store and load parameter (e.g,  a $\leq$ {(4,4): b}). 

These constraints can now be solved through bi-unification. Bi-unification eliminates constraints and produces unconstrained types by representing constraints on variables as intersection and union types \cite{mlsub}. A collection of lower bounds on a type can be represented as $\bigsqcup_ilb_i$ and upper bounds can be represented as $\bigsqcap_iub_i$.

\subsubsection{Bi-unification as Constraint Decomposition and Type Coalescing:}

MLSub uses bi-unification to produce unconstrained types from type constraints \cite{mlsub}. Bi-unification works by resolving all constraints to atomic constraints on variables and substituting atomic lower bounds for positive occurrences and atomic upper bounds for negative occurrences. SimpleSub presents a simple and efficient mutation based approach to bi-unification for producing unconstrained types \cite{simplsub}. The approach splits bi-unification into constraint decomposition and coalescing. BinSub takes this simplified approach to generating unconstrained types.

Decomposition collects constraints on type variables ensuring that all constraints implied by the initial constraint $\tau^+ \leq \tau^-$ and the typing rules of BinSub are represented by a consistent constraint set on type variables. A constraint set consists of upper and lower bounds for each type variable (corresponding to its negative and positive instantiations). A constraint set is consistent if all lower bounds of a type variable are subtypes of its upper bounds. A consistent constraint set is built from a subtyping constraint by deriving all constraints on variables implied by the typing rules for constructors involved in the component types, and then enforcing consistency by recursively applying rules to constrain the lower or upper bounds of a variable whenever a bound is added. The algorithm for constraint decomposition is highlighted in Algorithm \ref{alg:decomposition}. This algorithm extends SimpleSub with the corresponding pointer, record, and function rules for BinSub types. Rules for functions and records are omitted due to space, but follow the same pattern as pointers where \texttt{constrain} is called recursively on matching parameters depending on the covariance or contravariance of those parameters.

\begin{algorithm}[htbp]
\caption{Recursively decompose a constraint $\tau_0 \leq \tau_1$ into constraint set $C$}\label{alg:decomposition}
\begin{algorithmic}
\Function{constrain}{$C,\tau_0,\tau_1$}
\If{$\texttt{ptr}(a,b)=\tau_0 \And \texttt{ptr}(c,d)=\tau_1$}
    \State \Call{constrain}{$C,c,a$}
    \State \Call{constrain}{$C,b,d$}
\ElsIf{$ \alpha = \tau_0$} 
    \State \Call{concat}{$C[\alpha]_{ub}, \tau_1$}
    \ForAll{$x \in C[\alpha]_{lb}$}
        \State \Call{constrain}{$C,x,\tau_1$}
    \EndFor
\ElsIf{$ \alpha = \tau_1 $}
    \State \Call{concat}{$C[\alpha]_{lb}, \tau_0$}
    \ForAll{$x \in C[\alpha]_{ub}$}
        \State \Call{constrain}{$C,\tau_0,x$}
    \EndFor
\EndIf
\EndFunction
\end{algorithmic}
\end{algorithm}

\begin{figure}
    \centering
     \begin{allintypewriter}
        ; Assumes int32 is already in the lower bound of t3
        ; and d is already in the upper bound of e
        ; and t2 is in the upper bound of d
        constrain(t3, e)
            t3: {upper: [e], lower: [int32]}
            constrain(int32, e)
                e: {upper: [d], lower: [int32]}
                constrain(int32, d)
                    d: {upper: [t2], lower: [int32]}
                    constrain(int32, t2)
                        t2: {upper: [int32], lower: [int32]}
        ; Results
        stack_slot_1: {upper: [stack_slot_2})]
                            lower: []}
        stack_slot_2: {upper: [ ptr(a, {(4,4): b}), 
                                    ptr(c, {(0,4): d}),
                                    ptr({(0,4): e}, f)]})]
                            lower: []}
        t1: {upper: [stack_slot_1], lower: []}
    \end{allintypewriter}
    \caption{Interesting components of constraint decomposition on constraints from Figure \ref{fig:exampleconstraints}}
    \label{fig:constraintdecompexamp}
\end{figure}

Figure \ref{fig:constraintdecompexamp} shows interesting constraint decomposition calls on the example constraints from Figure \ref{fig:exampleconstraints}. The algorithm maintains a lower and upper bound on each type variable. When a type is constrained to be a subtype of another type (e.g., \texttt{constrain(t3,e)}) the bounds are modified to maintain consistent bounds. The figure shows calls to \texttt{constrain} (including recursive calls) and shows modifications to the upper and lower bounds of a variable during the call. The call \texttt{constrain(t3, e)} demonstrates maintaining consistency between bounds. When \texttt{e} is installed as an upper bound on \texttt{t3}, a lower bound of \texttt{int32} already exists on \texttt{t3}, so constrain must be called to apply the lower bound to \texttt{e} to maintain the requirement that  all lower bounds are subtypes of upper bounds. This new call to constrain then results on further constraint propagation to the variables \texttt{d} and \texttt{t2}.

Parreaux shows that this constraining process results in a consistent constraint set for the involved variables and that coalescing two types constrained by \texttt{constrain} result in two unconstrained types that are subtypes of one another \cite{simplsub}. This statement is the core lemma in proving the soundness of SimpleSub type inference. Coalescing a type $\tau$ uses polarity to produce a compact unconstrained representation of a type. The process of coalescing can be thought of as substituting a variable's lower bounds for its positive occurrences and a variable's upper bounds for its negative occurrences. This process is applied recursively to the bounds themselves with an occurrence check to capture recursive types. The rules for coalescing variables and pointers are highlighted in Algorithm \ref{alg:coalescing}. 

\begin{algorithm}[htbp]
\caption{Coalesce a type $\tau$ in the constraint context $C$ and polarity $p$ and recursive variables set $R$}\label{alg:coalescing}
\begin{algorithmic}
\Function{coalesce}{$C,\tau, p, R$}
\If{$\texttt{ptr}(a,b)=\tau$}
    \State \Return{\texttt{ptr}$($\Call{coalesce}{$C,a, \neg p, R$}$,$ \Call{coalesce}{$C,b, p, R$} $)$}
\ElsIf{$ \alpha = \tau$} 
    \If{$\alpha^p \in R$}
        \State \Return{$a$}
    \EndIf

    \State $S \gets \alpha$
    \State $bounds = C[\alpha]_{\textbf{if } p \textbf{ then } lb \textbf{ else } ub}$
    \State $rec = \Call{occurs}{\alpha^p, bounds}$
    \State $R' = R \cup \{ \alpha^p \}$
    \ForAll{$x \in bounds$ }
        \State $y=\Call{coalesce}{C,x,p, R'}$
        \If{p}
            \State $S \gets S \sqcup y$
        \Else
            \State $S \gets S \sqcap y$
        \EndIf
    \EndFor
    \If{$rec$}
        \State \Return{$\mu\alpha.S$}
    \Else
        \State \Return{$S$}
    \EndIf
\EndIf
\EndFunction
\end{algorithmic}
\end{algorithm}

As an example, coalescing \texttt{stack\_slot\_1} negatively begins by examining the upper bound 
\texttt{stack\_slot\_2}. Referring to the results in Figure \ref{fig:constraintdecompexamp}, the occurs check does produce a recursive type, because \texttt{stack\_slot\_2}'s upper bound refers to \texttt{b} which refers to \texttt{t1} which refers to \texttt{stack\_slot\_1}. Therefore, the variable representing \texttt{stack\_slot\_1}: $\alpha$ becomes recursively bound. Coalescing then proceeds to \texttt{stack\_slot\_2}. At each variable, the variable is coalesced with its upper bounds, using intersection types to take the least upper bound of all upper bounds. A negative position (e.g., store position) will cause the polarity to invert causing lower bounds to be aggregated by union types.  The unsimplified coalesced type for \texttt{stack\_slot\_1} is: 
\begin{multline*}
   \mu \alpha.\alpha \sqcap \texttt{stack\_slot\_2} \sqcap
   \texttt{ptr}(a, \{(4,4): b \sqcap (\textit{t1} \sqcap \alpha)\}) \sqcap \\ \texttt{ptr}(c, \{(0,4): d \sqcap (\textit{t2}  \sqcap \texttt{int32})\}) \sqcap \texttt{ptr}(\{(0,4): e \sqcup \texttt{int32}, f) 
\end{multline*}

This coalesced type offers a finite representation of the constraint set for \\ \texttt{stack\_slot\_1}, but contains redundancies. The following section shows how this type is optimized by representing coalesced types as a regular language \cite{algebraicdolan}. Simplification takes advantage of this representation to build type automata and apply well known automata minimization techniques to optimize the represented type \cite{noord2000treatment,hopcroft1971n}. 

\subsubsection{Type Simplification:}\label{sec:simplification}

BinSub uses automata based simplification inspired by MLSub rather than the co-occurrence based simplification presented in SimpleSub \cite{mlsub,simplsub}. It is unclear if the co-occurrence analyses presented capture the extent of desirable simplifications on type variables. Additionally, operating over recursive types directly instead of on automata leads to a complex recursive type merging procedure that must unroll recursive types until they align \cite{simplsub}. Type automata optimization strategies can directly leverage standard algorithms for automata simplification \cite{mlsub}. Utilizing this construction does have a limitation. By merging function and record constructors this optimization technique assumes type equalities of the form $a \rightarrow b \sqcap c \rightarrow d = a \sqcup c \rightarrow b \sqcap d$. This typing rule is desirable to achieve compact types in BinSub, but other type systems may wish to forgo this rule, requiring changes to the optimization strategy. 

Type automata based optimization consists of three steps: automata construction, determinization, and minimization. Automata determinization and minimization are standard aside from the maintenance of a lattice element attached to each node. When determinization or minimization merges nodes, the lattice elements are merged according to their polarity. Below we highlight the automata construction and definition of the node lattice.  

Automata are constructed from coalesced polar types. A type automata is defined by a tuple $<Q, q_0, \delta, H>$ where $q_0 \in Q$ and $\delta \in \Sigma$. The alphabet $\Sigma$, contains labels for all type parameters of type constructors, that is $\Sigma = \{ \epsilon, \texttt{FnIn}(n), \texttt{FnOut}(n)$ $, \texttt{RecLabel}(n,m),$ $\texttt{StoreLabel}, \texttt{LoadLabel}\}$ where $n,m \in \mathbb{N}$. The transition table is constructed by the rules:

\begin{figure}[H]
\begin{minipage}{.5\linewidth}
  $q(\tau_0 (\sqcap/\sqcup) \tau_1) \xrightarrow{\epsilon} q(\tau_0)$

  $q(\tau_0 (\sqcap/\sqcup) \tau_1) \xrightarrow{\epsilon} q(\tau_1)$

  $q(\texttt{ptr}(\tau_0,\tau_1)) \xrightarrow{\texttt{StoreLabel}} q(\tau_0)$

  $q(\texttt{ptr}(\tau_0,\tau_1)) \xrightarrow{\texttt{LoadLabel}} q(\tau_1)$
  
  \end{minipage}%
  \begin{minipage}{.5\linewidth}
  $q(\{... (n,m): \tau ...\}) \xrightarrow{\texttt{RecLabel}(n,m)} q(\tau)$
  
  $q((... n: \tau ...) \rightarrow (...)) \xrightarrow{\texttt{FnIn}(n)} q(\tau)$
  
  $q((...) \rightarrow (... n: \tau ...)) \xrightarrow{\texttt{FnOut}(n)} q(\tau)$

  $q(\mu\alpha.\tau) \xrightarrow{\epsilon} q(\tau)$
  \end{minipage}
\end{figure}

Each state is labeled with a lattice element and polarity described by the function $H$. The function \texttt{polarity} describes the polarity of a type in the current context. $V$ is the set of all type variables and $A$ defines an atomic type lattice.

\begin{figure}[H]
    $H(\tau) = (\texttt{polarity}(\tau),E(\tau))$
    
    $E: \tau \rightarrow L$
    
    $L: (T: \mathcal{P}(V) \times I: \mathcal{P}(\mathbb{N}) \times O:\mathcal{P}(\mathbb{N}) \times R: \mathcal{P}(\mathbb{N} \times \mathbb{N}) \times  P: \mathcal{P}(\{p\}) \times A) $
    
    $E(\alpha) = (\{\alpha\} ,\emptyset ,\emptyset, \emptyset, \emptyset,  \bot_A)$

    $E(\texttt{ptr}(\tau_0, \tau_1)) = (\emptyset ,\emptyset ,\emptyset, \emptyset, \{ p \}, \bot_A)$

    $E(\{ ... (n,m): \tau ... \}) = (\emptyset,\emptyset ,\emptyset, \texttt{dom}(\{ ... (n,m): \tau ... \}), \emptyset, \bot_A)$
    
    $E((... n: \tau,... ) \rightarrow (... m: \tau,... )) = (\emptyset,\texttt{dom}((... n: \tau,... )) ,\texttt{dom}((... m: \tau,... )), \emptyset, \emptyset, \bot_A)$

    $E(\rho) = (\emptyset ,\emptyset ,\emptyset, \emptyset, \emptyset,  \rho)$
\end{figure}

BinSub extends $L$ from prior work to incorporate lattices of in parameters ($I$), out parameters ($O$), record capabilities ($R$), and pointer capabilities ($P$) \cite{algebraicdolan}. The elements of $L$ form a product lattice of powerset lattices (excluding the atomic type lattice) so we define a meet and join operator accordingly (note the inverted direction of the lattice causing the join of two types to be a supertype of those types):

\begin{figure}[H]
$\alpha \sqcup \beta = (..,\alpha_i  \cap \beta_i,...,\alpha_\rho \sqcup_A \beta_\rho)$

$\alpha \sqcap \beta = (..,\alpha_i  \cup \beta_i,...,\alpha_\rho \sqcap_A \beta_\rho)$

\end{figure}

From this lattice definition we define a commutative partial function for merging nodes:
\begin{figure}[H]
$ M((\text{tt},l_0),(\text{tt}, l_1)) = (\text{tt}, l_0 \sqcup l_1)$

$ M((\text{ff},l_0),(\text{ff},l_1)) = (\text{ff}, l_0 \sqcap l_1)$

\end{figure}

Type optimization occurs by performing subset determinization on the type automaton and then performing DFA minimization. $M$ is used to merge node labels both during determinization and minimization. $M$ is sufficient to handle merging node labels during determinization due to the construction of the automaton. By definition, nodes are only reachable by $\epsilon$ if the types labeling the nodes have the same polarity. Similarly, nodes reachable by the same constructor will have the same polarity because the edge label implies the polarity of a destination node with respect to the source node. During minimization, the initial state partition is determined both by the polarity and the constructors in the lattice element of the node, preventing merging nodes of different polarity during the application of Hopcroft's minimization algorithm.

Figure \ref{fig:simpleaut} shows the simplified automata for \texttt{stack\_slot\_1}. Importantly, determinization reduces redundancies in type variables. The variables \texttt{d} and \texttt{t2} are merged into the same state due to epsilon edges between them in the original automata. The shown automata excludes load and store labels that only contain a single variable. This simplified type allows for a compact lowering.

\begin{figure}[htbp]
    \centering
    \begin{tikzpicture}[
       node distance=60pt,auto
    ]
    \node[state] (q_0)  {\scriptsize$(\{\alpha, \textit{t1}, b,...\}, \{p\})^-$};
    \node[state] (str) [below left= 30pt and 30pt of q_0] {\scriptsize$(\{(0,4)\})^+$};
    \node[state] (strfld) [below of=str] {\scriptsize$(\{e,d,\textit{t2}\}, \texttt{int32})^+$};
    \node[state] (slabel) [below right=30pt and 40pt of q_0] {\scriptsize$(\{(0,4)\})^-$};
    \node[state] (ldr) [right=100pt of q_0] {\scriptsize$(\{(4,4)\})^-$};
    \node[state] (ldrrec0) [below of=slabel] {\scriptsize$(\{d,\textit{t2}\}, \texttt{int32})^-$};
    \path[->] 
    (q_0) edge  node {\scriptsize\texttt{StoreLabel}} (str)
    (q_0) edge[bend left]  node {\scriptsize\texttt{LoadLabel}} (ldr)
    (slabel) edge  node {\scriptsize\texttt{RecLabel(0,4)}} (ldrrec0)
    (q_0) edge  node[left] {\scriptsize\texttt{LoadLabel}} (slabel)
    (ldr) edge[bend left]  node[above] {\scriptsize\texttt{RecLabel(4,4)}} (q_0)
    (str) edge  node {\scriptsize\texttt{RecLabel(0,4)}} (strfld);
    \end{tikzpicture}
    \caption{Simplified automata for \texttt{stack\_slot\_1}}
    \label{fig:simpleaut}
\end{figure}

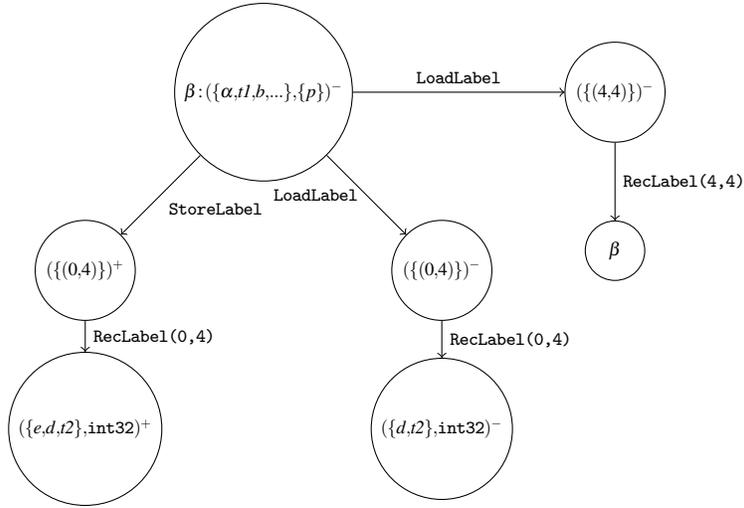
\begin{figure}[htbp]
    \centering
    \begin{tikzpicture}[
       node distance=60pt,auto
    ]
    \node[state] (q_0)  {\scriptsize$\beta: (\{\alpha, \textit{t1}, b,...\}, \{p\})^-$};
    \node[state] (str) [below left= 30pt and 30pt of q_0] {\scriptsize$(\{(0,4)\})^+$};
    \node[state] (strfld) [below of=str] {\scriptsize$(\{e,d,\textit{t2}\}, \texttt{int32})^+$};
    \node[state] (slabel) [below right=30pt and 30pt of q_0] {\scriptsize$(\{(0,4)\})^-$};
    \node[state] (ldr) [right=80pt of q_0] {\scriptsize$(\{(4,4)\})^-$};
    \node[state] (ldrrec0) [below of=slabel] {\scriptsize$(\{d,\textit{t2}\}, \texttt{int32})^-$};
    \node[state] (recnode) [below of=ldr] {\scriptsize$\beta$};
    \path[->] 
    (q_0) edge  node {\scriptsize\texttt{StoreLabel}} (str)
    (q_0) edge  node {\scriptsize\texttt{LoadLabel}} (ldr)
    (slabel) edge  node {\scriptsize\texttt{RecLabel(0,4)}} (ldrrec0)
    (q_0) edge  node[left] {\scriptsize\texttt{LoadLabel}} (slabel)
    (ldr) edge  node {\scriptsize\texttt{RecLabel(4,4)}} (recnode)
    (str) edge  node {\scriptsize\texttt{RecLabel(0,4)}} (strfld);
    \end{tikzpicture}
    \caption{Automata with loops broken}
    \label{fig:simpleautnoloops}
\end{figure}

\subsection{Type Lowering}\label{sec:typelowering}

Instead of decompiling automata back to BinSub types, BinSub produces C types from type automata. C types cannot represent type variables or intersection types so the lowering process is heuristic. There are two major algorithms required for building C types from type automata. The first algorithm selects how to break loops in the type automaton by creating named recursive types. This pass inserts synthetic named type nodes that consume internal strongly connected component edges until loops are broken. BinSub uses a heuristic to first consume edges that reach a pointer node from a structure field as the recursive variable. In Figure \ref{fig:simpleaut}, this heuristic results in a named node replacing the node $(\{\alpha, \textit{t1}, b,...\}, \{p\})^-$ for the edge: \texttt{RecLabel(4,4)}. During lowering, this name will refer to the resulting type for the named node. Figure \ref{fig:simpleautnoloops} shows the acyclic version of the type automata, naming the pointer $\beta$.

The second algorithm visits each node in the acyclic graph recursively and picks a type for that node. The type is selected from the lattice elements assigned to the node in a heuristically selected a-priori established priority: 1. record 2. pointer 3. function 4. atom. The selected constructor informs which outgoing edges will be used, and the polarity of the destination nodes informs whether to apply meet or join to repeated edges (with a caveat for pointers described below). In Figure \ref{fig:simpleautnoloops}, the root node receives the type pointer causing a meet on the targets of the load label resulting in $\{(0,4): \texttt{int32}\} \sqcap \{(4,4): \beta\}$. The store parameter is aggregated by performing a join of all store labels resulting in $\{(0,4): \texttt{int32}\}$. C types do not represent a separate load and store type parameter for pointers or intersection types so BinSub must merge these types. BinSub applies the following identities to merge multiple record or pointer labels into a single type:

\begin{gather*}
\{ (a,b): c\} \sqcap \{ (d,e): f\} = \begin{cases}
\{ (a,b): c \sqcap f\} &\text{$a = d \wedge b = e$}\\
\{ (a,b): c, (d,e): f\}  &\text{otherwise}
\end{cases}\\
\texttt{ptr}(a,b) \sqcap \texttt{ptr}(c,d) = \texttt{ptr}((a \sqcup c) \sqcap (b \sqcap d),b \sqcap d)\\
(a: b) \rightarrow (c: d) \sqcap (e: f) \rightarrow (g: h) = \begin{cases}
(a: b \sqcup f) \rightarrow (c: d \sqcup h) &\text{$a = e \wedge c = g$}\\
(a: b \sqcup f) \rightarrow (c: d, g: h) &\text{$a = e \wedge c \neq g$}\\
(a: b, e: f) \rightarrow (c: d \sqcup h) &\text{$a \neq e \wedge c = g$}\\
(a: b, e: f) \rightarrow (c: d, g: h)  &\text{otherwise}
\end{cases}
\end{gather*}

The pointer identity aggregates all load parameter types via a meet then performs a meet of the store parameters with the aggregate load parameter to enforce the constraint that store parameters are subtypes of load parameters for the \texttt{ptr} constructor. For the example type, the load parameters $\{(0,4): \texttt{int32}\} \sqcap \{(4,4): \beta\}$ are aggregated to: $\{(0,4): \texttt{int32}, (4,4): \beta\}$ using the record identity. Continuing the example, to compute the final pointer type, BinSub performs a meet of the store parameter with the aggregated load parameter, resulting in the named type: $\beta=\{(0,4): \texttt{int32}, (4,4): \beta\}$. This named type is the final type that can be rendered as the C type from Figure \ref{fig:linkedlist}. 

The definition of the record identity implies the possibility of overlapping fields. For instance, the type $\{(0,4): \tau_1\} \sqcap \{(2,2): \tau_2\}$ can be represented as $\{(0,4): \tau_1, (2,2): \tau_2\}$. BinSub's type system itself is non-normative in how such overlapping fields should be rendered as a C type and simply records the possible offset and size accesses for a given record. Lowering could produce a union for overlapping fields to mantain a sound C type that allows accesses at all offsets recognized by BinSub. Instead, BinSub currently aggregates smaller fields into larger fields as a heuristic to produce succinct types. 

\subsubsection{Polymorphism:}

Polymorphism has been ignored throughout the construction of the type inference and simplification algorithm. This focus on type inference and simplification internal to a function correlates with prior work \cite{retypd}. The lack of higher order functions and internal polymorphic bindings in C means polymorphism can be achieved at function boundaries through function cloning, rather than generalization during type inference. These assumptions are simplifying assumptions; function pointers cannot be treated polymorphically with this solution. Using this simplifying assumption we define a compositional inter-procedural polymorphic type inference algorithm, using the above definition of intra-procedural type solving. 

\begin{algorithm}[htp]
\caption{Infer a type automaton for each function in a binary for a list of SCCs S}\label{alg:infer}
\begin{algorithmic}
\Function{infer}{$S$}
    \State \Call{sort\_by}{$S, \texttt{reverse\_topo}$}
    \State $A \gets \{\}$
    \State $R \gets \{f \rightarrow \bot | f \in (s \in S)\}$
    \ForAll{$s \in S$}
        \ForAll{$(x, a) \in \Call{infer\_automata}{s,A}$}
            \State $A \gets A \cup  \{x \rightarrow a \}$
        \EndFor
        \ForAll{$c \in \texttt{callsites}(s)$}
            \State $R[c.\texttt{called\_function}] \gets R[c.\texttt{called\_function}] \sqcup \Call{infer\_automata}{c,A}$
        \EndFor
    \EndFor
    \State \Return{$R$}
\EndFunction
\end{algorithmic}
\end{algorithm}

Polymophism requires a procedure \texttt{decompile\_automata} that produces a BinSub type $\tau$ from an automaton. This procedure follows from prior work and forms an intersection or union type of all a node's type constructors, depending on the node's polarity. Recursive types are inserted when a cycle is detected \cite{algebraicdolan}. Using this procedure, global type inference proceeds backwards through the strongly connected components of the call graph. At function callsites, the type automata representing the function is decompiled to a BinSub type with fresh type variables, achieving polymorphism between callsites. After all functions' type automata are inferred, function signatures can be refined by applying join to all of the callsite types for a function. Joining type automata can be defined in terms of the optimization procedure defined above and forming a union of the two automata with epsilon edges. We highlight inter-procedural inference with refinement in Algorithm \ref{alg:infer}, referring to \texttt{infer\_automata} as the procedure for determining a type automaton for a function from its definition (by generating constraints) and type automata for callee functions. Strongly connected components of functions are treated together and are not polymorphic when called within an SCC.  

\section{Formalization of BinSub}\label{sec:formalbinsub}

Throughout the description of BinSub there are apparent similarities between the type system and Retypd. We now formalize this connection by describing a translation from Retypd constraints to BinSub constraints. We demonstrate that Retypd derived type variables can be interpreted by a BinSub typing judgement and that subtyping judgements derived between Retypd type variables can equivalently be derived using BinSub's rules. The correctness of BinSub type inference and Retypd type inference with respect to each type system's subtyping rules demonstrates a correspondence between the type inference approaches. This correspondence along with the following empirical efficiency results demonstrate the value in BinSub's simplified type system as an alternative view on Retypd type inference. We use the notation $\leq_R$ and $\leq_B$, to distinguish between the subtyping judgement in Retypd and BinSub, respectively.

\begin{figure}[ht]

    \centering
    \begin{minipage}{.5\linewidth}
         \begin{prooftree}
        \AxiomC{$x : \texttt{ptr}(a,b)$}
        \AxiomC{$a \leq_B b$}
        \LeftLabel{Pointer-Load}
        \BinaryInfC{$x.\text{load} : b$}
    \end{prooftree}

    \begin{prooftree}
        \AxiomC{$x : \texttt{ptr}(a,b)$}
        \AxiomC{$a \leq_B b$}
        \LeftLabel{Pointer-Store}
        \BinaryInfC{$x.\text{store} : a$}
    \end{prooftree}
    \end{minipage}%
    \begin{minipage}{.5\linewidth}
          \begin{prooftree}
        \AxiomC{$x : \{ (k,N) : \tau\}$}
        \LeftLabel{Record}
        \UnaryInfC{$x.\sigma N@k : \tau$}
    \end{prooftree}

    \begin{prooftree}
        \AxiomC{$x : (L: \tau) \rightarrow ()$}
        \LeftLabel{Function In}
        \UnaryInfC{$x.in_L : \tau$}
    \end{prooftree}

        \begin{prooftree}
        \AxiomC{$x : () \rightarrow (L: \tau)$}
        \LeftLabel{Function Out}
        \UnaryInfC{$x.out_L : \tau$}
    \end{prooftree}
    \end{minipage}
  
    \caption{Typing rules for Retypd derived type variables embedded in the typing judgement $:$}
    \label{fig:translation}
\end{figure}

Translation from Retypd types to BinSub type constraints is described by the rules in Figure \ref{fig:translation}. Using the rules in Figure \ref{fig:translation}, each Retypd constraint $\tau_{r1} \leq_R \tau_{r2}$ can be represented equivalently in BinSub by the constraint $\alpha \leq_B \beta$ where $\tau_{r1} : \alpha$ and $\tau_{r2} : \beta$. Retypd types are represented by derived type variables where a type variable is followed by a list of capabilities in the form: $\alpha.\textit{capability}$. Consequently, Retypd contains an additional judgement $\text{Var}(\alpha.\textit{capability})$ to assert that $\alpha$ has the capability $\textit{capability}$. Retypd has five capabilities: $load$, $store$, $in_L$, $out_L$, and $\sigma N@k$. These capabilities correspond to pointer load, pointer store, in parameter, out parameter, and field capabilities. $load$, $out_L$, and $\sigma N@k$ are covariant capabilities, while $store$ and $in_L$ are contravariant capabilities \cite{retypd}. Figure \ref{fig:translation} shows the direct relationship between BinSub constructors and Retypd capabilities. 
 
\begin{figure}[ht]
    
    \centering
    \begin{minipage}{.5\linewidth}
           \begin{prooftree}
        \AxiomC{$\triangleleft \Gamma \vdash \gamma \leq_B \alpha$}
        \AxiomC{$\triangleleft \Gamma \vdash \beta \leq_B \delta$}
        \LeftLabel{Pointer-$\oplus$-$\ominus$}
        \BinaryInfC{$\Gamma \vdash \texttt{ptr}(\alpha, \beta) \leq_B \texttt{ptr}(\gamma, \delta)$}
    \end{prooftree}

    \begin{prooftree}
        \AxiomC{$\triangleleft \Gamma \vdash \gamma \leq_B \alpha$}
        \AxiomC{$\triangleleft \Gamma \vdash \beta \leq_B \delta$}
        \LeftLabel{Function-$\oplus$-$\ominus$}
        \BinaryInfC{$\Gamma \vdash (L:  \alpha) \rightarrow (L: \beta) \leq_B  (L:  \gamma) \rightarrow (L: \delta)$}
    \end{prooftree} 
    \end{minipage}%
    \begin{minipage}{.5\linewidth}
        
    \begin{prooftree}
        \AxiomC{$\Gamma \vdash \alpha \leq_B \beta$}
        \AxiomC{$\Gamma \vdash \beta \leq_B \delta$}
        \LeftLabel{Transitivity}
        \BinaryInfC{$\Gamma \vdash \alpha \leq_B \delta$}
    \end{prooftree}
    \end{minipage}
    
    \caption{New typing rules for BinSub/Rules required for the proof sketch}
    \label{fig:typerules}
\end{figure}

The rules for BinSub's subtyping judgements are derived from Parreaux's rules in SimpleSub \cite{simplsub}. In Figure \ref{fig:typerules}, we highlight the new rules that are relevant for comparison with Retypd. These rules in combination with the simplification rules presented in Section \ref{sec:typelowering} encompass the extensions to rules beyond prior work. Rules for recursive types, records, and intersection and union subtyping remain the same. We restrict the function rule to a single label for simplicity. The later modality $\triangleleft$ introduced in the premises for the function and pointer constructor rules
 maintains the guardedness restriction required in SimpleSub's hypothetical rules \cite{simplsub}.
 
\subsubsection{Pointer Representation with Store and Load Variables:} \label{sec:pointerrep}

Using the rules from Figure \ref{fig:translation} and \ref{fig:typerules}, we can demonstrate how BinSub arrives at derivations of subtyping relationships in both covariant and contravariant cases of pointer subtyping. We use the example sets of constraints from Retypd \cite{retypd}: $\{q \leq_R p, x \leq_R p.\text{store}, q.\text{load} \leq_R y\}$ and $\{q \leq_R p, x \leq_R q.\text{store}, p.\text{load} \leq_R y\}$. Proving that $x \leq_R y$ in the first constraint set requires contravariance between the store parameter for $p$ and store parameter for $q$. Proving $x \leq_R y$ in the second constraint set requires covariance between the load parameter for $q$ and the load parameter for $p$. First we show that we can derive $x \leq_B y$ in the first constraint set using contravariance on store parameters, then we show a similar derivation for the second constraint set using covariance on load parameters. Applying the Pointer-Load rule from Figure \ref{fig:translation} to $q.\text{load}$ results in the type $\texttt{ptr}(q_s, q_l)$ and constraint $q_s \leq_B q_l$ and applying Pointer-Store to $p.\text{store}$ results in the type $\texttt{ptr}(p_s, p_l)$ and constraint $p_s \leq_B p_l$. Substituting these types into the constraint set yields the BinSub constraints (leaving $x$ and $y$ as type variables): $\{p_s \leq_B p_l, q_s \leq_B q_l, \texttt{ptr}(q_s, q_l) \leq_B \texttt{ptr}(p_s, p_l), x \leq_B p_s, q_l \leq_B y\}$. We recognize that Pointer-$\oplus$-$\ominus$ from Figure \ref{fig:typerules} is invertible and thus when applied to $\texttt{ptr}(q_s, q_l) \leq_B \texttt{ptr}(p_s, p_l)$ implies the constraints: $\{p_s \leq_B q_s, q_l \leq_B p_l \}$. A pointer subtyping judgement of the form $\texttt{ptr}(q_s, q_l) \leq_B \texttt{ptr}(p_s, p_l)$ can only arise through this rule and transitivity and union and intersection types will preserve these constraints on parameters to the pointer constructor. This inversion lemma is derived in  Dolan and Mycroft \cite{mlsub}.  Combining the constraints $x \leq_B p_s$, $p_s \leq_B q_s$, $q_s \leq_B q_l$, and $q_l \leq_B y$ by transitivity yields the expected $x \leq_B y$. In the covariant direction, the starting constraint set transforms into the BinSub set: $\{p_s \leq_B p_l, q_s \leq_B q_l, \texttt{ptr}(q_s, q_l) \leq_B \texttt{ptr}(p_s, p_l), x \leq_B q_s, p_l \leq y\}$. Using the same pointer rule, we can use the covariant constraint: $q_l \leq_B p_l$, and combine this constraint with the constraints $x \leq_B q_s$, $q_s \leq_B q_l$, and $p_l \leq y$ to derive $x \leq y$. These derivations show BinSub can derive judgements with covariant loads and contravariant stores. 

\subsubsection{Relationship of Rules to Retypd:}

\begin{figure}[ht]
    
    \centering
    \begin{minipage}{.5\linewidth}
    \begin{prooftree}
        \AxiomC{$\alpha \leq_B \beta$}
        \AxiomC{$\text{Var}(\beta.l)$}
        \AxiomC{$<l>=\oplus$}
        \LeftLabel{S-Field$_\oplus$}
        \TrinaryInfC{$\alpha.l \leq_B \beta.l$}
    \end{prooftree}

    \begin{prooftree}
        \AxiomC{$\alpha \leq_B \beta$}
        \AxiomC{$\text{Var}(\beta.l)$}
        \AxiomC{$<l>=\ominus$}
        \LeftLabel{S-Field$_\ominus$}
        \TrinaryInfC{$\beta.l \leq_B \alpha.l$}
    \end{prooftree}
    \end{minipage}%
    \begin{minipage}{.5\linewidth}
        \begin{prooftree}
        \AxiomC{$\alpha \leq_R \beta$}
        \AxiomC{$\beta \leq_R \gamma$}
        \LeftLabel{S-Trans}
        \BinaryInfC{$\alpha \leq_R \gamma$}
    \end{prooftree}    
    
    \end{minipage}

    \caption{Relevant Retypd subtyping rules for comparing Retypd to BinSub}
    \label{fig:retypdrules}
\end{figure}

To describe the relationship between BinSub and Retypd subtyping rules we have included the relevant rules from Retypd in Figure \ref{fig:retypdrules}. These rules and the S-Pointer rule shown in Figure \ref{fig:spointerrule} are the rules applied in a normal proof tree by Retypd \cite{retypd}. The rule S-Field$_\oplus$ applies to covariant capabilities while the rule S-Field$_\ominus$ applies to contravariant capabilities. The rules have the expected behavior of placing the types represented by the capability as a subtype of each other in the direction that corresponds to the capability's variance. 

The relationship highlighted in Section \ref{sec:pointerrep} can now be formalized in the claim that any Retypd derivation between two derived type variables can be represented by a BinSub derivation between the BinSub types: $C \vdash x_R \leq_R y_R \implies \forall x_B,y_B.(x_R : x_B \wedge y_R : y_B \implies C' \vdash x_b \leq y_b)$ where $C'=\{ x_B \leq_B y_B | x_R: x_B  \wedge y_R: y_B \wedge x_R \leq_R y_R \in C \}$ We now provide a sketch of the proof for this claim. We first rely on a lemma proved by Retypd: for any derivation in Retypd there exists a normal proof that does not use the rules: T-Prefix, T-Inherit, or S-Refl. Retypd also considers a constraint set closed under T-Left and T-Right \cite{retypd}. Therefore we need only show corresponding derivations to proof trees that contain the rules S-Trans, S-Pointer, $\text{S-Field}_\oplus$, and $\text{S-Field}_\ominus$. For S-Trans, it can be shown that the corresponding transitivity rule shown in Figure \ref{fig:typerules} in BinSub applies for any pair of Retypd subtyping constraints related by S-Trans. The relationship between transitivity rules is direct. For covariant and contravariant labels, BinSub's typing rules reify these rules directly into the rules for type constructors. To show the correspondence between BinSub and Retypd label rules, we consider a label in relation to it's associated BinSub constructor. We take as an example a derivation with the label $out_L$ where $C \vdash \alpha.out_L \leq_R \beta.out_L$ by the covariant field rule for $out$. In such a case, $\alpha \leq_R \beta$ and $\text{Var}(\beta.out_L)$ are both in $C$. By the definition of $:$ and $C'$ we then have $a: x$, $b: y$,  $x \leq_B y$ and $y=(...) \rightarrow (L: \tau_1)$ in $C'$. From these premises we show that $\forall \tau_2.( a.out_L : \tau_2 \implies \tau_2 \leq_B \tau_1)$, demonstrating a derivation that corresponds to Retypd's covariant field rule for out fields. First by inversion of the typing rules defined by $:$ (valid by the uniqueness of rules for constructors), it must be the case $a: (...) \rightarrow (T: \tau) \wedge \tau = \tau_2$, so we can substitute this type for $x$ and the corresponding type for $y$ into $x \leq_B y$ to get $(...) \rightarrow (T: \tau_2) \leq (...) \rightarrow (L: \tau_1)$. By the inversion lemma described above (used to prove type soundness in MLSub), we can invert the rule Function-$\oplus$-$\ominus$ to derive $\tau_2 \leq_B \tau_1$, proving the desired subtyping judgement. Symmetric reasoning suffices to prove this correspondence for the covariant field rule of $load$ and records as well as the contravariant rules for $in_L$ and $store$. We now provide a sketch of the correspondence proof for the S-Pointer rule from Figure \ref{fig:spointerrule}. Given that Var $\alpha.load$ and Var $\alpha.store$ are in $C$ we must prove that $\forall x,y.( \alpha.store : x \wedge \alpha.load : y \implies x \leq_B y )$. This lemma holds by the inversion of $:$. Inverting Pointer-Store from Figure \ref{fig:translation} with respect to $\alpha.store : x$ gives $\alpha : \texttt{ptr(x,y)} \wedge x \leq_B y$ and inversion of Pointer-Load with respect to  $\alpha.load : y$ gives the same type of $\alpha$. From the premises of the rule we have $x \leq_b y$ for $\alpha.store : x$ and $\alpha.load : y$, proving the desired subtyping judgement.

This correspondence between derivations in Retypd and BinSub demonstrates that type inference performed in BinSub can derive the same expressive subtyping judgements as Retypd. BinSub's efficient type inference and simplification algorithm therefore provide a compelling alternative to Retypd's weighted pushdown system.

\section{ Evaluation }
\subsection{Implementation in Angr} \label{angrimpl}
To empirically evaluate the efficiency and precision of BinSub, we implemented the type constraint decomposition, coalescing, simplification, and lowering algorithms of BinSub in Angr \cite{wang2017angr}. Angr contains an implementation of Retypd constraint generation in its variable recovery pass. Typehoon, Angr's type inference algorithm, implements type simplification using the weighted pushdown system from Retypd. This combination of features makes comparison between BinSub and Retypd in Angr natural. We implemented a type inference algorithm that leverages BinSub to simplify Retypd constraints and produce lowered types. The BinSub type inference pass uses the rules defined in Figure \ref{fig:translation} to derive BinSub constraints from Angr's Retypd constraints. By maintaining identical interfaces, we can directly compare BinSub to Typehoon by running both passes as part of Angr's decompilation process and comparing the resulting types to ground truth. Both Typehoon and BinSub are able to use the heuristics provided by Angr's intermediate language brightening passes prior to constraint generation, and use the same atomic type lattice.   

\subsection{Evaluation Setup}

To compare Typehoon (an implementation of Retypd) and BinSub, we ran both type simplification algorithms on a dataset sampled from the ALLSTAR reverse engineering dataset \cite{allstarDataset}. To construct our dataset, we start from 200 randomly selected binaries from the amd64 and armel categories. During the evaluation, we perform type inference on sampled functions within a binary in sequence. We repeat each function benchmark 7 times with a 1 minute timeout. We run each binary for a global timeout of 2800 seconds and record samples for which all 7 benchmarks terminated without timing out, prior to the 2800 second binary-level timeout. For each function in a binary, we run both type inference algorithms as part of Angr's Clinic pass (the collection of pre-decompilation analyses and optimizations) and record the inferred type signature for the function. These timeouts make evaluation tractable. The binary-level timeout ensures that large binaries from the sample do not dominate the dataset, and the function-level timeout ensures we make progress within the 22 hour evaluation window. The total number of functions in the original set of binaries is 135646 with a median number of 114 functions per binary. On average each binary receives 7 comparable functions within the 2800 seconds spent per binary (400 seconds per microbenchmark). These comparable functions result in a dataset of 1,568 functions for which both BinSub and Typehoon terminate within the one minute function timeout (with only 43
function timeouts in BinSub and 161 function timeouts in Typehoon with 19 overlapping failing functions)\footnote{The mean function byte size in the 19 overlapping failed functions is 1083 and the median byte size is 888. These statistics show a 518\% increase in mean function size for failing functions when compared with the overall statistics for comparable functions in Table \ref{tab:evalfunccomplexity}. This increase in average size indicates expected behavior where larger functions are reaching the timeout.}.  We name the new dataset of comparable functions ALLSTAR\_1568. The constraint complexity of functions in these binaries is recorded in Table \ref{tab:evalfunccomplexity}. The constraint complexity is proportional to the number of instructions for a function as constraint generation generates a constant number of constraints per instruction type. For these 1568 functions that successfully receive a signature from both BinSub and Typehoon, we record the time spent in the type recovery algorithm and the distance between the inferred signature and the ground truth signature from DWARF information \cite{bastian2019reliable}. The mean run time for all seven benchmarks is used as the overall time for that function. All data is presented with respect to comparable samples, because timing information is not available for functions that timed out.

\subsubsection{Type Distance Metric:}
The distance metric is inspired by TIE, but makes some simplifications to handle record padding and arbitrary numbers of return values \cite{tiebap}. The distance between function types is computed as the average of the distance between both the parameter types and return types. Any label for a parameter or return value that is not in both the ground type and inferred type is assumed to be equal to the maximum distance in the type lattice. The distance for records is computed similarly as the average distance of all matching labeled types with any non-matching labels assumed to have the maximum lattice distance. There is a caveat for records: labeled fields in the inferred type that do not overlap with ground type fields are filtered prior to comparison. This filtering prevents considering recovered padding during comparison between the ground type and inferred type. Finally, atomic types' distances are defined by the average distance of each atomic type to the lowest common ancestor of the atomic types in $A$. Any type that is incomparable is assumed to have a distance equal to the maximum distance in the subtyping lattice. This distance metric captures an intuitive measure of the precision of a subtyping-based binary type inference algorithm as the lattice distance between the ground and recovered type. 

\begin{table}
  \label{tab:evalfunccomplexity}
  \caption{Complexity of Comparable Functions from ALLSTAR\_1568}
  \centering
\begin{minipage}{.4\linewidth}
    \begin{tabular}{|c|c|}
    \hline
    Metric & Value (Constraints) \\
    \hline
    Max & 704 \\
    Median & 18 \\
    Mean & 35 \\ 
    \hline
    \end{tabular}
\end{minipage}%
\begin{minipage}{.4\linewidth}
\begin{tabular}{|c|c|}
    \hline
    Metric & Value (Bytes) \\
    \hline
    Max & 3260 \\
    Median & 96 \\
    Mean &  175 \\ 
    \hline
    \end{tabular}

\end{minipage}
\end{table}

\subsubsection{Variation in Benchmarks:} To validate the precision of our micro-benchmarks we recorded timing benchmarks for 1000 functions of the comparable dataset and computed the standard deviation for the runtime of each sample as the percentage of the mean. The average mean standard deviation was 3\% of the mean for that sample.

\subsection{Results}

Table \ref{tab:evalres} shows the average type distance and average runtime for type inference on the 1568 comparable functions in the dataset. As expected, BinSub and Typehoon have similar precision, differing by .01 in type distance from the DWARF ground truth when measured by the TIE distance algorithm defined above in "Type Distance Metric" \cite{tiebap}. On the other hand, BinSub does not need to saturate a pushdown system and explore a constraint graph in order to infer types (the time intensive operations in Retypd) leading to a 63x reduction in mean runtime when comparing the average runtime per function between the two analyses. We also computed the percent reduction in time per function and computed a 95\% confidence interval on the mean percentage improvement, arriving at the interval: (87\%,88\%), providing strong statistical evidence for BinSub's efficiency. This confidence interval was computed by bootstrapping the distribution with an iteration count of 75,000. Empirical comparison is important for validating the performance of BinSub. While BinSub does not require path exploration of the constraint graph or the saturation algorithm from Retypd, the powerset construction used for determinization can create $2^N$ DFA states. This empirical demonstration shows that, in practice, types in binaries result in NFAs where determinization is tractable.

\begin{table}
  \caption{Evaluation Results on Comparable  ALLSTAR\_1568 Functions}
  \label{tab:evalres}
  \centering
  \begin{tabular}{|c|c|c|}
    \hline
    Algorithm & Average Type Distance & Average Runtime (seconds/function)  \\
    \hline
    BinSub & 1.67 & .024 \\
    Typehoon & 1.68 & 1.51 \\
    \hline
    \end{tabular}
\end{table}

\begin{figure}[ht]
\begin{minipage}{.5\linewidth}
        \centering
        \includegraphics[width=150pt]{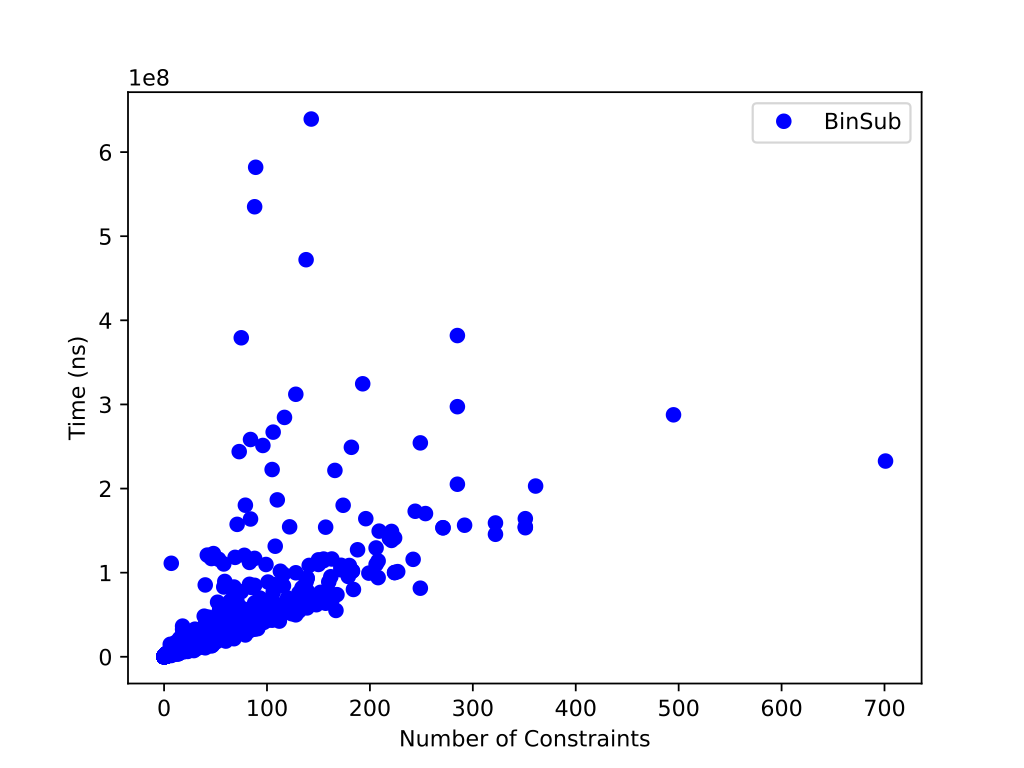}
        \caption{BinSub Time (1e8 ns) vs Size}
        \label{fig:binsubtimevsconstraintsize}
\end{minipage}%
\begin{minipage}{.5\linewidth}
  
        \centering
        \includegraphics[width=150pt]{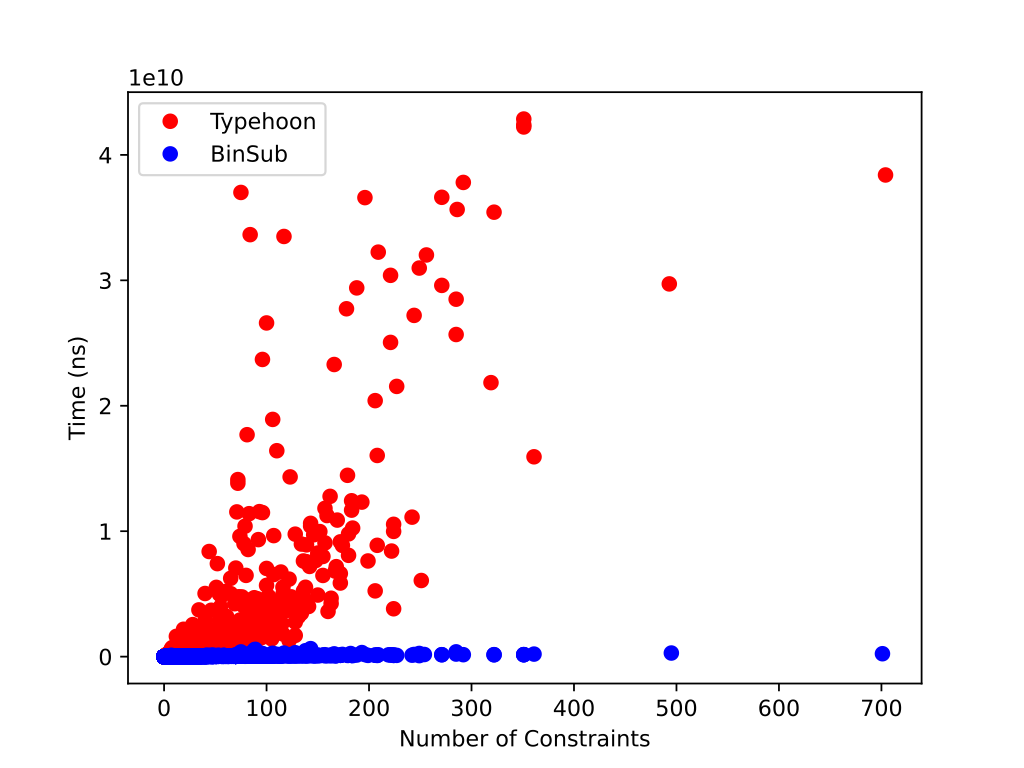}
        \caption{Time (1e10ns) vs Size}
        \label{fig:typehoontimevsconstraintsize}
    
\end{minipage}
\end{figure}

Figure \ref{fig:binsubtimevsconstraintsize} and \ref{fig:typehoontimevsconstraintsize} highlight the performance of each algorithm as the size of the constraint set grows. Retypd shows the expected rapid growth in runtime as the number of constraints increases, while BinSub remains fast on functions with a large number of constraints. Figure \ref{fig:binsubtimevsconstraintsize} presents a closer look (using a smaller time unit size of 1e8 ns) at BinSub's performance against the number of constraints. The structure of some functions and types causes increases in constraining and optimization time, but on average BinSub shows marked improvement in performance over Retypd. BinSub's simplicity lends itself to an efficient implementation that scales to complex functions. 
  
\section{Conclusions and Future Work}

BinSub demonstrates that critical requirements for a type system for binary type inference can be captured elegantly with algebraic subtyping. Subtyping, polymorphism, recursive types, and contravariant stores with covariant loads can all be represented in the framework of algebraic subtyping without a weighted pushdown automata. The simplification of binary type inference to algebraic subtyping lends itself to a scalable implementation of precise type inference. Perhaps more importantly, this reduction to algebraic subtyping places binary type inference within a lineage of work in traditional type inference. BinSub unifies binary type inference with current work in type inference for high-level languages. BinSub can benefit from future work on principal type inference for ML with subtyping based on algebraic subtyping.

BinSub's implementation could benefit from further investigation along these lines. The type simplification strategy is modular and likely could be tailored to produce simpler C types from type automata by extending BinSub's simplification rules with further equivalences. Additionally, the current type lowering heuristics do not leverage precise reasoning rules to distinguish pointers from integers at a given use site and simply prioritize pointer types over atomic types. Integration of Retypd's pointer rules could increase BinSub's precision in distinguishing pointers from integers during lowering \cite{retypd}. BinSub's initial implementation offers a baseline capability on top of which further incremental precision improvements can be made in type simplification and lowering. 

\begin{credits}
\subsubsection{\ackname} This work was performed at Trail of Bits and we would like to thank the internal reviewers: Michael Brown, Artem Dinaburg, and Josiah Dykstra in providing feedback that greatly improved the paper's presentation. Peter Goodman also provided invaluable questions that helped refine the ideas presented in type simplification. Additionally, we would like to thank the shepherd who helped refine this paper in a number of ways. We would also like to thank the anonymous reviewers for their invaluable time and thoughtful critique. 

\end{credits}
\bibliographystyle{splncs04}
\bibliography{sample}

\end{document}